\documentclass{nic-series}

\def\lsi{\raise0.3ex\hbox{$<$\kern-0.75em\raise-1.1ex\hbox{$\sim$}}}
\def\gsi{\raise0.3ex\hbox{$>$\kern-0.75em\raise-1.1ex\hbox{$\sim$}}}

\newcommand{\nn}{\nonumber}

\newcommand{\la}{\langle}

\newcommand{\gsim}{\mathop{\gsi}}
\newcommand{\R}{{\kern+.25em\sf{R}\kern-.78em\sf{I} 
  \kern+.78em\kern-.25em}}
\newcommand{\C}{{\kern+.25em\sf{C}\kern-.50em\sf{I} \kern+.50em\kern-.25em}}

\begin{document} 

\title{
Extracting physics from an unphysical situation:\\
\vspace*{3mm}
light mesons in a small box}

\author{W. Bietenholz\inst{\, 1}, 
S. Capitani\inst{\, 1}, 
T. Chiarappa\inst{\, 2}, 
M. Hasenbusch\inst{\, 2},
K. Jansen\inst{\, 2}, \\
M. M\"uller-Preussker\inst{\, 1},
K.-I. Nagai\inst{\, 2},
M. Papinutto\inst{\, 3},
S. Shcheredin\inst{\, 1}, \\
A. Shindler\inst{\, 2},
C. Urbach\inst{\, 2,}  \inst{\, 4} and
I. Wetzorke\inst{\, 2}}

\institute{Humboldt Universit\"{a}t zu Berlin, Institut f\"{u}r Physik\\
Newtonstr.\ 15, D-12489 Berlin, Germany \\  \  \\
         \and  
         NIC/DESY Zeuthen, \\
         Platanenallee 6, D-15738 Zeuthen, Germany\\  \  \\
         \and
         DESY Hamburg, \\
         Notkestr. 85, D-22603 Hamburg, Germany\\  \  \\
         \and
         Insitut f\"{u}r Theoretische Physik,
         Freie Universit\"{a}t Berlin\\
         Arnimallee 14, D-14195 Berlin, Germany
          }

\maketitle

\begin{abstracts}
Quantum Chromo Dynamics is considered in a setup where
the light mesons are squeezed into unphysically small boxes. We show
how such a situation can be used to determine the couplings of the low
energy chiral Lagrangian from lattice simulations, applying chirally 
invariant formulations of lattice fermions. \\

{\bf Preprint \ \ DESY 03-128, \ HU-EP-03/57, \ SFB/CPP-03-32}

\end{abstracts}

\section{Introduction}

A straightforward approach for 
numerical simulations in statistical mechanics or in high energy
physics is to perform them in boxes that are large enough for 
the relevant correlation lengths --- or the Compton wave lengths
of the lightest particles --- to live comfortably inside the box.
This is an ideal world and, as it might have been suspected,
as far as numerical simulations are concerned this world 
is also very expensive.
Approaching the critical point of a second order phase transition, 
where also a {\em continuum} field theory is defined, 
one needs to increase the correlation length, accompanied by the 
corresponding extension
of the box size. For models in high energy physics the number of lattice
points needed in the simulations has to be increased 
with the fourth power in this limit, 
and this does not even include additional factors originating from the 
scaling behavior of the algorithms employed. 

It is one of the fascinating discoveries that physical models 
of interest can be considered
in {\em unphysical} situations while it is still possible to extract  
correct physical information from them. The reason is that characteristic 
properties of a model do not change when it is considered under 
unphysical conditions. 
The advantage of this idea is that often the model 
can be considered in a setup 
where numerical simulations are much easier than in the case  
of a large physical volume.

Let us give a --- not too serious --- example to illustrate this idea:
think of a folding chair. If it is unfolded (its physical state), 
you recognize it easily as a chair and you can use it as such.
However, it would not fit into your car. Folding it (the unphysical state), 
you can put it in the
car, but it is not recognizable easily as a chair anymore. Still, it is a
chair, of course, and by measuring the length of the struts etc.\ (which 
do not change in the folded state) you can deduce the size of the 
original chair. 
 
The idea of studying a system as a function of the box size 
in order to extract physical information is quite old 
and proved to be very fruitful. 
Finite size scaling arguments were already introduced 1883 by
Reynolds for turbulence studies in air and liquid flows.
Later on, they were applied in investigations of critical phenomena
at phase transitions to extract critical exponents \cite{critical},
to determine scattering lengths \cite{scat} and to renormalize
scale dependent quantities such as the running 
strong coupling constant, the running quark masses and 
so-called renormalization constants
\cite{martinreview,rainerreview}. 
The physical problem we want to consider in the present article is 
the dynamics
of Goldstone bosons as they appear in the case of a  
spontaneously broken continuous symmetry, such as the $O(N)$ symmetry 
in non-linear $\sigma$-models
and the chiral symmetry in Quantum Chromo Dynamics (QCD)
with massless quarks.

The dynamics of these Goldstone bosons can be described by
{\em chiral perturbation theory} \cite{gale}, which evaluates the 
chiral Lagrangian. This Lagrangian describes the low energy 
properties of some underlying, more fundamental theory. It is constructed 
such that it obeys the same (global) symmetries as the fundamental theory. 
Examples are
effective descriptions of the $\Phi^4$-theory and, of course,
QCD for which chiral perturbation theory was designed.
The structure of the chiral Lagrangian is very general.
In leading order it can  be written as
\be
{\cal L}_{\rm eff}[U] = \frac{F_{\pi}^{2}}{4} {\rm Tr}
\Big[ \partial_{\mu} U \partial_{\mu} U^{\dagger} \Big] -
\frac{1}{2} \Sigma m_{q} {\rm Tr} \Big[ U e^{i\theta /N_{f}}
+ U^{\dagger} e^{-i\theta /N_{f}} \Big] \ .
\label{chiraL}
\ee
Here $U(x) \in SU(N_{f})$ represents the Goldstone boson field,
$N_{f}$ is the number of flavors, $m_q$ is the quark mass
\footnote{For simplicity we assume the same quark mass
for all flavors. In a spin model it corresponds to an
external magnetic field. Strictly speaking, due to $V < \infty$ and
$m_q > 0$ this kind of symmetry breaking is not fully spontaneous
and we actually deal with remnants of the Goldstone bosons.}
and $\theta$ is the vacuum angle.
The Lagrangian in eq.~(\ref{chiraL}) contains two so-called low
energy constants (LEC), the {\em pion decay constant} $F_\pi$ and the order 
parameter of chiral symmetry breaking, i.e.\ the {\em scalar condensate}
$\Sigma$. 
Chiral perturbation theory allows for a systematic higher order 
expansion with more complicated terms in the Lagrangian and corresponding
additional LEC multiplying these terms. 
The LEC are free parameters of the Lagrangian. They
can only be determined by the comparison with sources beyond
chiral perturbation theory. In principle, their values follow from
the underlying fundamental theory, in this case QCD. 

In the infinite volume (the standard regime of chiral 
perturbation theory)
the Goldstone field $U$ is parameterized as 
$U(x) = \exp\left\{i\sqrt{2}\xi(x)/F_{\pi} \right\}$
with a fluctuating field $\xi(x)$. When the volume is taken to be small,
the Compton wavelength of the Goldstone boson exceeds the finite
physical extent 
$L$ of the box. In this situation, the so-called {\em $\epsilon$-regime}
\cite{xpt,hansen}, the
constant mode $U_0$ is separated and the field is written as 
$U(x) = U_0\exp\left\{i\sqrt{2}\bar \xi (x)/F_{\pi} \right\}$ 
with $\int_{V} \bar \xi (x) \, dx =0$. 
The contribution of the constant mode 
$U_0$ to the chiral Lagrangian must be treated non-perturbatively. 

\begin{figure}[t]
\begin{center}
\epsfxsize=8cm
\epsfysize=8cm
\epsfbox{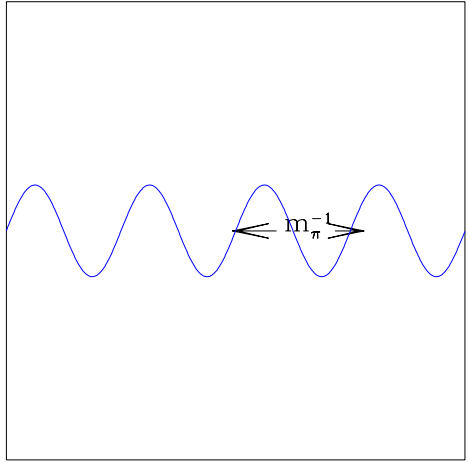}
\vspace{-0.7cm}
\caption{\label{figure0}
Illustration of the $p$-expansion regime. The Compton wavelength of the 
pion fits into the finite box.  
}
\end{center}
\vspace{-0.2cm}
\end{figure}

\begin{figure}[t]
\begin{center}
\epsfxsize=8cm
\epsfysize=8cm
\epsfbox{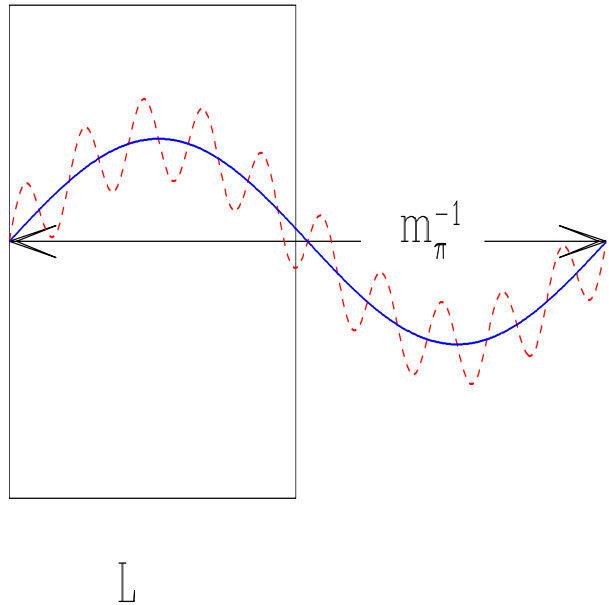}
\caption{\label{figure0a}
Illustration of the $\epsilon$-regime. The Compton wavelength
(blue solid line) exceeds the length of the box, whereas the fluctuations 
(red dashed line) do fit into the box.}
\end{center}
\vspace{-0.2cm}
\end{figure}

The situation is illustrated in Figure \ref{figure0} for the 
$p$-expansion and in Figure \ref{figure0a} for the $\epsilon$-expansion.
In the $p$-expansion regime, the pion Compton wavelength fits into the
finite box, corresponding to an infinite volume situation. 
In the  $\epsilon$-regime, the pion Compton wavelength exceeds
the box length and the zero momentum pion mode appears as a constant
mode. Notice, however, that the fluctuations around the constant mode 
are still well accommodated within the box.

Numerical simulations are a powerful tool
to determine the LEC. These simulations use  
the fundamental Lagrangian of QCD itself. Tuning 
the quark masses such that the regime of chiral perturbation 
theory is reached, a comparison of numerically generated QCD data 
and the predictions of chiral perturbation theory can be performed. 
In this way the LEC can be extracted by a fit of
the numerical data to the analytical formulae of chiral perturbation theory, 
which contain the LEC as free parameters. 
The so determined values of the LEC as originating from QCD can 
then be related to experimental data leading to direct tests of QCD. 
In addition, the simulations can --- in principle --- 
be performed exploring different scenarios.  
For example one of the quark masses can be put to zero \cite{mu0}
and it can be tested whether such a scenario would be consistent 
with the real world. 

Knowing the LEC is also an important ingredient to supplement 
further numerical simulations 
themselves. Simulations at small values of the quark mass,
corresponding to their physical values, are very expensive, see e.g.
Ref.\ \cite{panel2001}. If we would know, however, 
at what values of the quark mass chiral perturbation theory is valid,
chiral perturbation theory itself could be used to 
extrapolate many observables to physical values of the quark mass. 
Obviously, for this procedure the knowledge of the LEC would be
crucial. 
 
Now the obvious question arises why such comparisons of 
simulation results and chiral perturbation theory have not been done
before. The answer is, {\em they have}, but the interpretation of the 
outcome has been difficult and no convincing picture emerged.
For the infinite volume simulations it is still not clear when contact 
to chiral perturbation theory can be established. Various 
additional assumptions have to be incorporated into the analytical 
computations in chiral perturbation theory 
to be able to relate them to numerical results \cite{panel2002}.

The finite volume simulations, which seem to be much easier, were hampered
by the fact that the lattice formulations of fermion actions either 
break chiral symmetry explicitly (Wilson fermions) or that topological
sectors are hard to distinguish (staggered fermions) \cite{staggered}. 
The identification of topological charge sectors is a necessary prerequisite
for exploring the $\epsilon$-regime, since here 
the observables depend strongly on the topology \cite{LeuSmi}. 
Similar problems are encountered
for improved versions of these lattice fermions. 

A great leap forward was achieved with the rediscovery 
\cite{Hasenfratz:1998ri} of the 
Ginsparg and Wilson~\cite{chiral:GWR} relation, which reads for some
(yet to be specified) lattice Dirac operator $D$ at $m_{q}=0$
\begin{equation}
\gamma_5D + D\gamma_5 = a D\gamma_5D\; .
\label{gwrelation}
\end{equation}
Clearly, in the limit of a vanishing lattice spacing $a$
the usual anti-commutation relation of the continuum 
Dirac operator is recovered. 
The Ginsparg-Wilson relation implies an exact lattice chiral symmetry 
\cite{Martin_exact}
if the action is constructed with a lattice Dirac operator that solves
the Ginsparg-Wilson relation. Consequently Ginsparg-Wilson fermions
have a well defined fermionic index. By means of the Index Theorem, 
this property also allows for a conceptually clean separation of 
topological sectors \cite{Hasenfratz:1998ri}. Thus one
overcomes the obstacles that other formulations of 
lattice fermions are plagued with. 

For completeness we
give a particular example for a solution of the Ginsparg-Wilson 
relation found by 
H.~Neuberger \cite{chiral:ovlp} from the overlap formalism \cite{ovlp},
based on the pioneering work by D. Kaplan \cite{kaplan}. 
To this end, we first consider the standard 
Wilson-Dirac operator on the lattice,
\begin{equation}
  D_{\rm w}=\frac{1}{2}\left\{\gamma_{\mu}(\nabla_{\mu}^{*}+\nabla_{\mu})
  -a\nabla_{\mu}^{*}\nabla_{\mu}\right\} \ ,
\label{dw}
\end{equation}
with $\nabla_{\mu}$, $\nabla_{\mu}^{*}$  
the lattice forward resp.\  backward derivatives, i.e.\ 
nearest neighbor differences acting on a field $\Phi(x)$,
\begin{eqnarray}
\nabla_{\mu} \Phi (x) &=& 
\frac{1}{a} \Big[ U(x,\mu ) \Phi (x + a \hat \mu) - \Phi (x) \Big] \ , \\
\nabla^{*}_{\mu} \Phi (x) &=& 
\frac{1}{a} \Big[ \Phi (x) - U ( x - a \hat \mu, \mu)^{\dagger} 
\Phi (x-a \hat \mu) \Big] \ . \nn
\end{eqnarray}
Here $U(x,\mu )$ is the link variable pointing from site $x$ into
the direction $\mu$, and $\hat \mu$ is a unit vector in the same 
direction. We then define 
\begin{equation}
A = D_{\rm w} - 1 - s \ ,
\end{equation} 
where the parameter $s$ can be tuned in some interval. 
At last Neuberger's overlap operator
$D_{\rm N}$ with mass $m$ is given by
\begin{eqnarray}
D_{\rm N} &=& \left\{ 1 - \frac{m_{q}}{2(1+ s)} D_{\rm N}^{(0)} \right\} 
+ m_{q} \ , \\
D_{\rm N}^{(0)} &=& (1+s) \left[ 1 + A(A^{\dagger}A)^{-1/2} \right]\; . \nn
\end{eqnarray}

Despite the appearance of the square root which connects all the lattice 
point with each other, the operator is local (in the field theoretical sense) 
as long as the gauge coupling is not too strong \cite{hjl}. 
However, the numerical 
implementation of the square root operator is very demanding
and restricts present simulation to the quenched approximation
\footnote{In this approximation virtual quark anti-quark states 
are completely neglected. Although this seems to be a very crude
approximation, it works surprisingly well in practice \cite{quench}.} 
, see also the reviews \cite{nieder,pilar}.
A promising approach for a construction of improved overlap
operators is to replace $D_{\rm w}$ in eq.~(\ref{dw}) by alternative
operators as proposed in Refs.\ \cite{hypercube}.

\section{Random Matrix Theory}

Let us now start our discussion of finite volume physics with the 
example of a special technique called Random Matrix Theory (RMT).
In many complex systems eigenvalues and their correlations    
play an important r\^{o}le. These eigenvalues may exhibit 
universal properties that 
can be described by RMT for many physical systems \cite{RMTbooks}.
Among the numerous application fields of RMT, also 
the low lying eigenvalues in the
QCD spectrum are expected to be described by RMT
(for a review, see Ref.\ \cite{tilormt}).
The theoretical background for this expectation is the fact that at 
zeroth order of chiral perturbation theory in the 
$\epsilon$-regime, taking only the constant mode 
$U_0$ into account, the Lagrangians of chiral perturbation theory 
and the one of RMT are equivalent.
Correspondingly, the LEC of chiral perturbation theory 
enter also the predictions of RMT, which in turn allows for their 
determination by confronting numerical data from lattice
QCD simulations with the theoretical formulae from RMT. 

Such a comparison of the predictions of RMT with numerical simulations is,
however, difficult again, because also in this case the knowledge of 
the topological charge sectors and an exact chiral symmetry are important. 
Both difficulties can be overcome elegantly with the use of 
operators solving the Ginsparg-Wilson relation. The limitation in this
case will only be the computer time available, but with the machines 
available at NIC such a project becomes feasible, though with a
rather limited statistics. 

Another difficulty is that the finite volume cannot be arbitrarily small. 
Note that the Goldstone bosons of the chiral symmetry
breaking pick up some mass if a small quark mass is switched on.
The crucial point of chiral perturbation theory is that these 
quasi-Goldstone bosons represent the light mesons.
Therefore, this effective low energy description can only work
in the world of mesons as bound states.
Hence the physical volume should --- roughly speaking ---
be larger than the confinement scale.
As we will see below, RMT can provide a quantitative answer
to the question about the scale where the validity of
chiral perturbation theory sets in.

\begin{figure}[t]
\begin{center}
\includegraphics[angle=90,angle=90,angle=90,width=12.0cm,height=10.0cm]{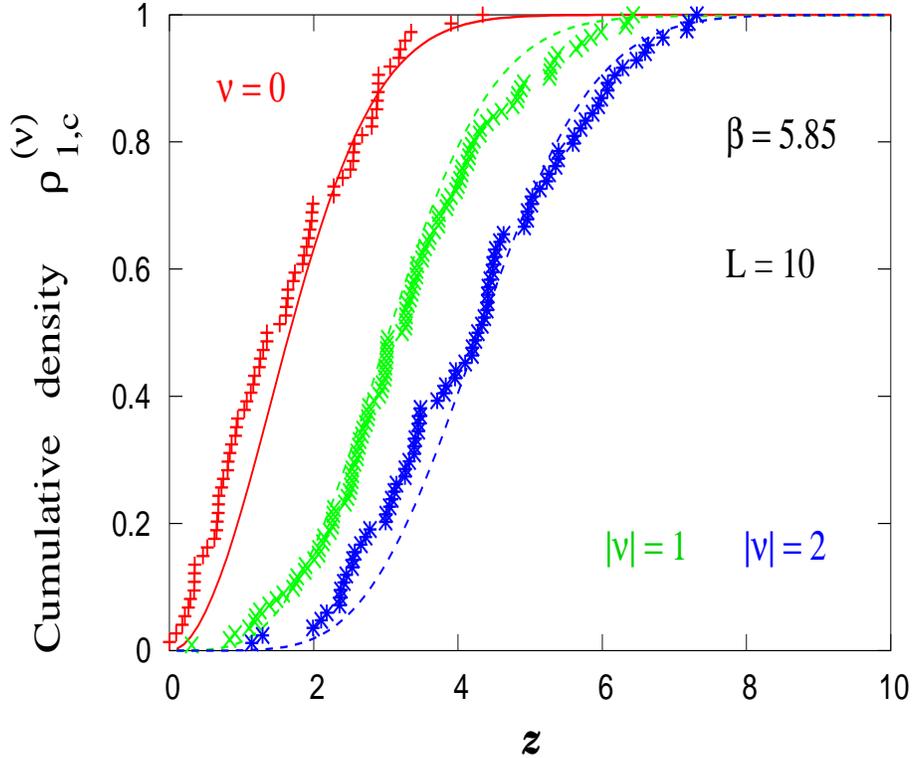}
\caption{\label{figure1}
The cumulative probability distribution for the first non-zero
eigenvalue in the topological sectors with charge 
$\vert \nu \vert =0,1,2$, 
on a lattice of size $V = (1.23 ~ {\rm fm})^{4}$.}
\end{center}
\end{figure}

In Figure~\ref{figure1} we show a result of such a computation
\cite{stanislav}. It addresses the cumulative 
probability distributions (see Ref.~\cite{numrec}, Chapter 14) of the
lowest (non-zero) Dirac eigenvalue $\lambda_{1}$.
For these distributions the predictions
of RMT in various topological sectors (solid lines)
are compared with the numerical data of our
simulations using the overlap operator. We see that the data from the 
quenched simulations are well described by RMT. 

Some remarks are in order. The first is that 
the data agree with the theoretical predictions
only if the lattice 
corresponds to a physical volume of $V \gsim (1.2 {\rm ~ fm})^4$. 
Going below this size, the predictions collapse.
Hence a minimal box length of about $1.2~{\rm fm}$ is necessary 
to be in the mesoscopic world where chiral perturbation and RMT 
work. The second remark is that from the probability distribution 
a value of the scalar condensate can be extracted and it is found 
to be consistent with earlier simulation results.
The third remark concerns higher eigenvalues.
Here the agreement is not as good as in the case of the leading
non-zero eigenvalue. Generally the RMT predictions are confirmed
up to some value of the dimensionless parameter $z = \lambda \Sigma V$.
This threshold raises gradually if the volume increases, and it
might be related to the so-called Thouless energy \cite{tilormt}.

A last remark is of a more general nature: the prediction of the 
eigenvalue distribution by RMT for the lowest non-zero
eigenvalue \cite{lowEV2} in topological charge sector zero reads
$P(\lambda)= \frac{z}{2}\exp (-\frac{1}{4}z^2 )$.
Hence, for simulations in this sector we expect to encounter quite 
frequently very small eigenvalues,
which will contribute in quark propagators as
$1/\lambda$. Now, from our comparison with RMT we know
that the RMT predictions are well respected by the 
numerical data. Hence the small eigenvalues in topological 
charge sector zero {\em have to appear} in the simulations with a
non-negligible probability.
Clearly, they will give rise to 
substantial fluctuations in physical observables, providing  
exceptionally large contributions proportional to the inverse of
the eigenvalue.

Of course, when the quark mass is chosen large enough it will act as an 
infrared regulator and therefore cut off the effects of these 
very small modes.
In the $\epsilon$-regime, however, we want to study the system 
at {\em small} quark masses. This leads to the problem of finding a window
for simulations in the $\epsilon$-regime: if the quark mass is too 
small, it cannot act as a regulator anymore and the small modes will
spoil the statistical sample. When, on the other hand, the quark is chosen
to be too large, we leave the $\epsilon$-regime. 
The situation is clearly better when we choose a topologically 
non-trivial sector. Here RMT predicts an eigenvalue  distribution 
that suppresses low eigenvalues substantially, rendering the simulations
much safer. 
The problem discussed here is of a very general nature and does also 
apply to the case of dynamical fermions. Therefore, it has to be expected that
these simulations become very demanding and problematic if performed
close to physical values of the quark masses. This is another motivation
to understand the contact with chiral perturbation very well, in order 
to let chiral perturbation theory do the job of computing observables
at the physical point. 

\section{Meson correlation functions}

In a previous NIC proceedings contribution \cite{nicproc}
we reported about a test
of spontaneous chiral symmetry breaking in QCD.
For that study the evaluation of the zeroth order of chiral 
perturbation theory (involving only $U_0$) was necessary. 
In this follow-up project we also computed meson correlation functions 
in the $\epsilon$-expansion regime for which the next order 
was known in the full theory \cite{hansen}, 
but not in the quenched approximation.  Therefore, new analytical 
computations were necessary for the 
quenched situation, a work that is published in Refs.\ \cite{qCPT,vecax}.
On the numerical side, the necessary propagators were computed
using the overlap operator.  

Restricting ourselves first to computations in topological charge sector 
zero, we found the correlation functions to be very noisy and with
our limited statistics 
it was not possible 
to extract a conclusive signal. We checked that this phenomenon can be
understood from the eigenvalue distribution of RMT. 
Focusing on the contribution of the lowest eigenvalue alone
(which is the largest contribution) and
following Ref.\ \cite{cond:paperI} we estimated
the statistics required to compute the scalar condensate
from the correlation function based on the 
eigenvalue distribution of RMT. Indeed, we found that 
$O(10^{4})$ configurations would be necessary to obtain reliable errors
\cite{KIN}.

Therefore, we did not explore the topological charge sector zero any further
and concentrated on topological charge sector one \cite{TC}. 
Repeating the analysis
from the theoretical eigenvalue distribution of RMT we found that with O(100)
configurations the errors tend to stabilize. 
From chiral perturbation theory we expect that in topological
charge sectors $\pm \nu$  the correlation function 
of the axial current takes the form (in a volume $L^{3} \times T$)
\bea  \label{axialcurrent}
\la A_{\mu}(0) \, A_{\mu}(t) \rangle_{\nu} &=&
\frac{F_{\pi}^{2}}{T}
\left[ 1 + \frac{2 m_{q} \Sigma_{\vert \nu \vert}(z_q ) 
T^{2}}{F_{\pi}^{2}} \cdot
h_{1}(\tau ) \right] \ , \\
h_{1}(\tau ) &=& \frac{1}{2} \Big( \tau^{2} - \tau + \frac{1}{6} \Big)
\ , \qquad \tau = \frac{t}{T} \ , \nn \\
\Sigma_{\nu}(z_q ) &=& \Sigma \left( z_q \Big[ I_{\nu}(z_q ) K_{\nu}(z_q ) +
I_{\nu +1}(z_q ) K_{\nu -1}(z_q ) \Big] + \frac{\nu}{z_q } \right)\ , 
z_q = m_q \Sigma V \nn \ .
\eea

\begin{figure}[t]
\begin{center}
\epsfxsize=12cm
\epsfysize=12cm
\epsfbox{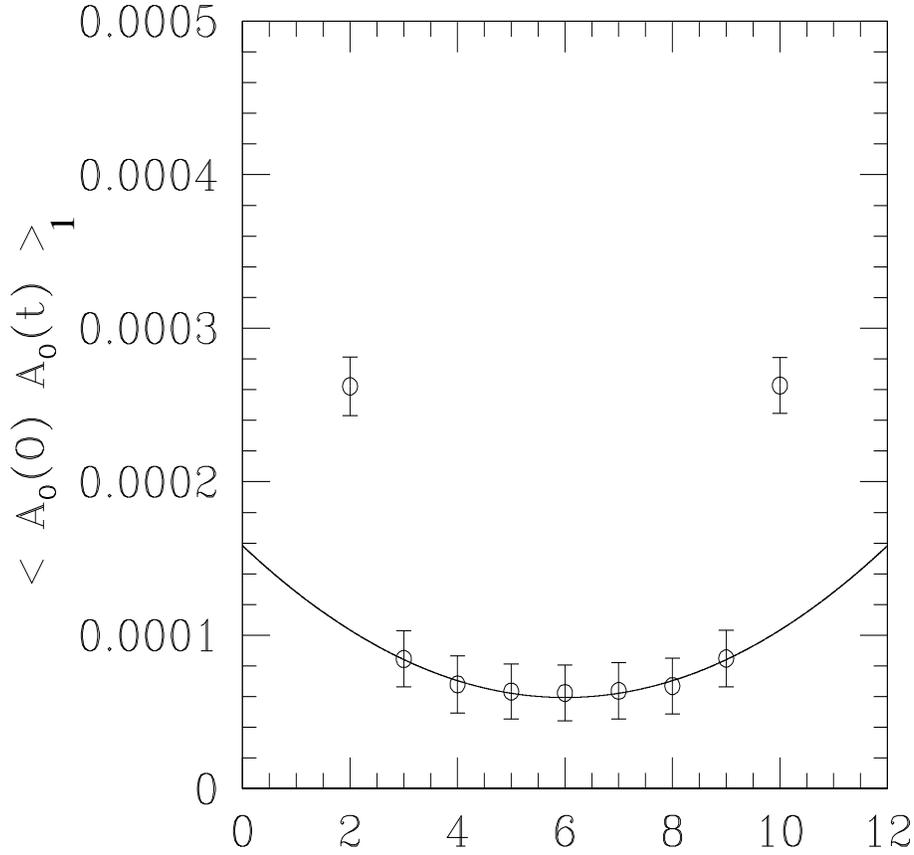}
\vspace{-0.7cm}
\caption{\label{figure3}
The axial correlation function on a $12^4$ lattice at 
$\beta=6$ in topological charge sector one. 
The solid line is a fit to the data using 
eq.~(\ref{axialcurrent}).
}
\end{center}
\vspace{-0.2cm}
\end{figure}

The first observation is that this correlation function does not 
show an exponential decay 
but a power law behavior, a clear reflection of the
fact that the pion Compton wavelength is larger than the box size. 
In the axial correlation function there appear again the two LEC 
of the effective Lagrangian (\ref{chiraL}), 
$F_{\pi}$ and $\Sigma$. 
In Figure~\ref{figure3} we show a fit of our data to the 
prediction of eq.~(\ref{axialcurrent}). For our simulations we used a $12^4$
lattice at $\beta=6$ and worked in topological charge sector one. 
From the fit, the value of $F_{\pi}$ can be determined quite reliably, 
whereas the value of $\Sigma$ is rather insensitive and cannot be extracted.
The situation here is somehow complementary to the
fit of the spectrum to the RMT predictions that we discussed in Section 2.

Nevertheless, we see that chiral perturbation theory can be used
to compute the LEC from meson correlation functions. The example of 
the axial current presented here can be extended to the scalar and 
the pseudo-scalar correlation functions (the vector correlation function
is identically zero \cite{vecax}), from which further LEC can be evaluated. 
However, in the formulae for those correlation functions additional
parameters show up.

\section{Conclusions}

The somewhat unconventional $\epsilon$-regime of chiral perturbation
theory turns out to be an interesting, but also difficult region to
be explored by means of numerical simulations. 
The parameters of the simulation 
have to be chosen with care: the topological charge must not be zero and
the value of the quark mass has to be in a certain window in order to 
avoid problems with small eigenvalues on one side, and to avoid 
leaving the $\epsilon$-expansion regime on the other side. If these
precautions are taken care of, the LEC of the chiral Lagrangian 
can be computed from the numerical simulations with powerful consequences
for future simulations in general.

The $\epsilon$-regime served in this study as
a kind of service setup to
provide the LEC from ``easy'' to be done numerical simulations.
By exploring the hypothetical world of quark masses that are much smaller
than the physical ones --- which can now be done in numerical simulations ---
the phenomenology of such a world could be tested. Possibly, in this
way we could learn why the quarks have the masses they assume in
nature. The results
that will be obtained in the $\epsilon$-regime might reveal
some answers to such questions in the future and hence this regime
may turn out to be not so unphysical after all. \\                               

{\bf Acknowledgment} \ \ 
This work was supported in part by the DFG Sonderforschungsbereich
Transregio 9, ``Computergest\"{u}tzte Theoretische 
Teilchenphysik''.

\end{document}